\documentclass[prb,aps, superscriptaddress, reprint, amsmath,amssymb, showpacs]{revtex4-1}
\usepackage{graphicx}
\usepackage{amsmath}
\usepackage{dcolumn}
\usepackage{bm}
\usepackage[%
colorlinks=true,
urlcolor=blue,
linkcolor=blue,
citecolor=blue
]{hyperref}
\usepackage{gensymb}
\begin{document}

\title { \textit{Ab initio} study on Quasi-Binary Acetonitriletriide Sr$_3$[C$_2$N]$_2$}
\author{ James Sifuna}
\email{jsifuna@cuea.edu}

\affiliation{ Department of Natural Science, 
	The Catholic University of Eastern Africa,
	62157 - 00200, Nairobi, Kenya.}
\affiliation{ Materials Modeling Group, 
	Department of Physics and Space Science,
	The Technical University of Kenya,
	52428-00200, Nairobi, Kenya.}  
\author{George S. Manyali}
\email{gmanyali@kafuco.ac.ke}
\affiliation{Computational and Theoretical Physics Group, Department of Physical sciences,
	Kaimosi Friends University College,
	385-50309, Kaimosi, Kenya.}
\author{ Elicah Wabululu}
\affiliation{ Department of Natural Science, 
	The Catholic University of Eastern Africa,
	62157 - 00200, Nairobi, Kenya.}
\affiliation{Department of Physics, 
	Kenyatta University,
	P.O Box: 43844-00100, Nairobi, Kenya.}   

\author{ Carolyne Songa}
\affiliation{ Department of Natural Science, 
	The Catholic University of Eastern Africa,
	62157 - 00200, Nairobi, Kenya.}
\author{ Alloysious Otieno}
\affiliation{ Department of Natural Science, 
	The Catholic University of Eastern Africa,
	62157 - 00200, Nairobi, Kenya.}
    
\author{ Stephen Sironik}
\affiliation{ Department of Natural Science, 
	The Catholic University of Eastern Africa,
	62157 - 00200, Nairobi, Kenya.}   

\date{\today}
\begin{abstract}
We report using density functional theory (DFT), the ground-state properties of the recently synthesized and characterized Sr$_3$[C$_2$N]$_2$ crystal. The nearly colorless, centrosymmetric Sr$_3$[C$_2$N]$_2$ crystallizes in a monoclinic unit cell with a $P2_1/c$ space group (No.14) and many of its properties remain unknown basing on the fact that it's a latecomer in the field. The goal of this study is to fill this information gap through a theoretical prediction. The calculated structural properties were comparable to those obtained by an experimental group led by Clark and co-workers thus giving us extra confidence in the accuracy of our DFT computations on Sr$_3$[C$_2$N]$_2$. We employed the same approach in calculating mechanical and dynamical stabilities together with the electronic density of states of Sr$_3$[C$_2$N]$_2$. No imaginary phonon modes were observed and thus implying dynamical stability. The thirteen elastic constants calculated passed the stability criteria of a monoclinic system. From the computed Poisson's ratio ($\eta$=0.27) and G/B=0.54, our calculations predict Sr$_3$[C$_2$N]$_2$ being brittle and not able to withstand high-pressure applications. To analyze the chemical bonding mechanism, the corresponding total density of states (TDOS) and partial DOS were plotted. The top of the valence band (VB) mainly consists of C 2p states N 2p, N 2s  and a slight admixture of Sr 5s states. The bottom of the conduction band (CB) shows a strong hybridization between C 2p, N 2p, N 2s, and  Sr 5s states, yielding a bandwidth of 7.18 eV in the entire conduction band. We were able to obtain a tunable electronic gap  of 2.65 eV in Sr$_3$[C$_2$N]$_2$. The authors herein note that Sr$_3$[C$_2$N]$_2$ falls in an unknown family of pseudonitrides that may possess novel physical and chemical properties if combined with suitable cations like transition metals. Material scientists are encouraged to scout in this new class of pseudonitrides for future technologies.

\end{abstract}


\maketitle

\section{Introduction}
\label{sec:Introduction}
Designing materials based on the first principles approach, synthesis, and equally characterization of these materials are of great interest to both theoretical and experimental material scientists\cite{james}. The search for new novel materials has become a norm in many material-based research groups around the world. The exponential search has been triggered by the enormous success in the current theoretical and experimental approaches. As an example, if carbon is combined with a light element like nitrogen, the outcome has always been promising from previous theoretical studies\cite{G1,G2,G3}. Unfortunately, few of these complex carbon-nitrogen-ions have been realized experimentally\cite{james}.\par
After the discovery of pseudo-chalcogenide anion [CN$_2$]$^{2-}$ in Ca[CN$_2$], there has been a continuous\cite{CN2} urge to obtain carbodiimide-based and cyanamides-based materials for potential applications. It is important to note that [CN$_2$]$^{2-}$ may exist as the symmetric carbodiimide anion or the asymmetric cyanamides depending on the hardness of the cation in place. Literature\cite{CN2,1994,2016,2001,2004} hints that carbodiimides like M[CN$_2$] (M=Mg - Ba and Eu), Cr$_2$[CN$_2$]$_3$ and M[CN$_2$] (M=Mn - Zn) as well as binary cyanamides such as M[M=Cd and Pb], are employed as negative electrode materials for lithium and sodium batteries, corrosion protective layers, photovoltaic devices, fluorescent light sources and light-emitting diodes. The carbodiimide anion is a pseudo-chalcogenide anion and can act as a bridging ligand to allow magnetic bridged paramagnetic cations like Cr$_2$[CN$_{2}$]$_3$\cite{2016}.
\par Most recently, a new unknown family of pseudonitrides was realized. The first quasi-binary acetonitriletriide Sr$_3$[C$_2$N]$_2$, was reported by Clark and co-workers\cite{paper}. They characterized Sr$_3$[C$_2$N]$_2$ by single-crystal X-ray diffraction, Raman spectroscopy, elemental analysis and confirmed it by quantum mechanical methods. Our study was promoted by the extensive work in Ref.~\onlinecite{paper} to aid fill the information gap on Sr$_3$[C$_2$N]$_2$. To this date, based on our knowledge, no studies whether theoretical or experimental have been performed on the mechanical stability and band-structure analysis of monoclinic Sr$_3$[C$_2$N]$_2$.
\par It is important to note that any material possesses its intrinsic characteristics. If these traits are well predicted and understood, then it becomes easy to manipulate or tailor a material for any novel functionality. Information regarding the lattice constants is usually important in the growth of thin layers on other materials. A mismatch can easily lead to strains and thus defects\cite{msc,thesis}. Mechanical stability on the other hand, is a very fundamental aspect of any material and it is based on its elastic constants. The elastic constants determine the way a crystal will respond to external forces and also give room to investigate stability, stiffness, brittleness, and ductility in a material. Knowledge of the band-structure contribution from the orbitals is important such that it depicts the chemical bonding nature in a crystal as explained in Ref~\onlinecite{msc}. Again, it can predict its optical capabilities if the gap seems tunable. With all these in mind, we filled the information gap on Sr$_3$[C$_2$N]$_2$ by performing an \textit{ab initio} study. We first bench-marked our study with the work of Clark and co-workers in Ref.~\onlinecite{paper} on lattice parameters and the optimized coordinates. Once satisfied, we went ahead to investigate the elastic constants, electronic density of states and the phonon bandstructure.
\par This paper is organized as follows in the remaining parts: in Sec.~\ref{sec:comps}, we give a brief outline of the calculation details. The results are shown and discussed in Sec.~\ref{sec:results}. Finally, in Sec.~\ref{conc} we give a conclusion and propose future works on monoclinic Sr$_3$[C$_2$N]$_2$.

\section{Calculation details}
\label{sec:comps}
 \begin{figure}[h!]
	\begin{center}
		\includegraphics[width=0.4\textwidth]{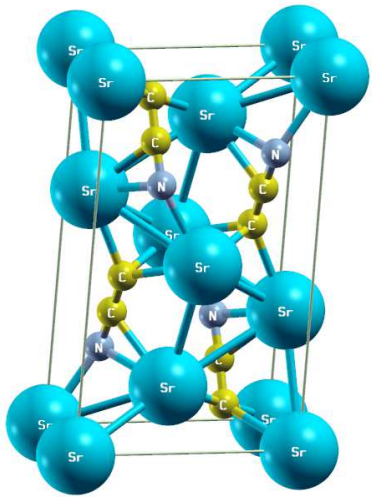}
		\caption{ We display the conventional cell of monoclinic Sr$_3$[C$_2$N]$_2$ in $P2_1/c$ space group (No.14) considered in the present calculation. The blue, green and gray balls represent Sr, C and N respectively. Information regarding the optimized atomic coordinates (fractional) has been given in Table \ref{positions} to ensure reproducible calculations in future. Other important aspects are found in Table \ref{table:0cn}.} 
		\label{fig:equi}
	\end{center}
\end{figure}
We performed scalar-relativistic calculations on Sr$_3$[C$_2$N]$_2$ using density functional formalism as implemented in the {\sc {Siesta}} method\cite{siesta} with a double zeta polarized basis set. Exchange and correlation were treated within the generalized gradient approximation (GGA) using the PBE\cite{pbe} functional. Core electrons were replaced by \textit{ab initio} norm-conserving fully separable \cite{KB}, Troullier-Martin pseudopotentials \cite{TroulMartin}. In {\sc{Siesta}}, the one-electron eigenstates were expanded in a set of numerical atomic orbitals using the standard {\sc Siesta} DZP basis. A Fermi-Dirac distribution with a temperature of 0.075 eV was used to smear the occupancy of the one-particle electronic eigenstates. To get a converged system, we had a two-step procedure: We first relaxed the atomic structure and the one-particle density matrix with a sensible number of k-points (9$\times$5$\times$7 Monkhorst-Pack\cite {monkhorst, soler_kpts} k-point mesh) and secondly, freezing the relaxed structure and density matrix, we performed a non-self consistent band structure calculation with a much denser sampling of 60$\times$60$\times$60. Real-space integration was carried over a uniform grid with an equivalent plane-wave cutoff of 600 Ry. In this calculation, all the atomic coordinates were relaxed until the forces were smaller than $0.01$ eV/\AA ~and the stress tensor components were below 0.0001 eV/\AA$^3$.

\begin{table}
	\caption{ Fractional coordinates of the studied Sr$_3$[C$_2$N]$_2$ in the monoclinic phase}
\label{positions}
\begin{tabular}{c c}
	
	\hline 
Atom	& Position (x,y,z) \\ 
	\hline 
Sr&0.000000000         0.000000000         0.000000000\\
Sr&0.000000004         0.499999854         0.500000328\\
Sr&0.678485771         0.152045862         0.435470093\\
Sr&0.321513943         0.847953807         0.564530748\\
Sr&0.321518608         0.652042412         0.064527652\\
Sr&0.678482154         0.347957366         0.935472290\\
N&0.087449979         0.210397309         0.200261007\\
N&0.912551701         0.789606001         0.799739143\\
N&0.912523899         0.710388851         0.299719372\\
N&0.087475085         0.289613275         0.700280840\\
C&0.235772471         0.184132419         0.721581733\\
C&0.764227685         0.815867535         0.278418430\\
C&0.764246370         0.684131708         0.778427357\\
C&0.235751809         0.315867237         0.221572747\\
C&0.384750637         0.075614761         0.736019744\\
C&0.615256520         0.924379726         0.263979027\\
C&0.615257998         0.575610654         0.763983118\\
C&0.384737027         0.424384036         0.236016889\\
	\hline 
\end{tabular} 
\end{table}

\section{RESULTS AND DISCUSSION}
\label{sec:results}
In this section we present the calculated ground-state properties of monoclinic Sr$_3$[C$_2$N]$_2$ as predicted using the PBE\cite{pbe} functional. It is important to note that we fitted our energy-volume relationship to the Murnaghan\cite{munaghan} equation of state. 
\subsection{Structural properties}

\label{sec:structural}
\begin{table*}[tb]
	\centering
	\caption{ Structural parameters of Sr$_3$[C$_2$N]$_2$: $a$ (\AA), $b$ (\AA), $c$ (\AA), $\alpha (\degree)$, $\beta (\degree)$, $\gamma (\degree)$, volume (\AA $^3$),  C-C bond (\AA), C-N bond (\AA) and the density (g/cm$^3$).}
	\label{table:0cn}
	\begin{tabular*}{\textwidth}{@{\extracolsep{\fill}}llllllllllllllllll@{}}
		\hline\noalign{\smallskip}
		Reference &a &b &c &$\alpha$& $\beta$&$\gamma$&volume&C-C&C-N&$\rho$\\ 
		\noalign{\smallskip}\hline\noalign{\smallskip}
		This work (Z=2) [DFT]&4.0552 &10.8702&7.0128 &89.9999&103.246&90.001&300.90&1.318&1.287&3.741\\
		Experiment\cite{paper}&4.0745&10.7254&7.0254&- &102.700&-&299.50&1.291&1.271&3.758\\
		\noalign{\smallskip}\hline
	\end{tabular*}
	\label{str}
\end{table*}
From the calculated parameters in Table \ref{str}, it is seen that the volume of Sr$_3$[C$_2$N]$_2$ crystal is 300.90 \AA$^3$ which is in agreement with the recent experimental value of 299.50  \AA$^3$. Our volume is 0.47\% higher than that of Clark and co-workers\cite{paper} due to slight distortion in our optimized lattice constants. The unit cell slightly increased by 1.35\% in the $b$ axis while we had a decrease of 0.47\% and 0.12\% in $a$ and $c$ axis respectively. This indicated an existence in different bonding characters along the three axes. The C-C and C-N bonds were overestimated as compared to the experimental values by 2.09\% and 1.26\% respectively.  Such a discrepancy is a well-known factor since PBE functional overestimates the lattices and thus the bonds.  Our calculated volumetric density slightly decreased by 0.45\% and this was attributed to the increased volume after optimization. The $\beta$ angle in our calculations was 0.53\%  larger than that in Ref.~\onlinecite{paper} and it is attributed to the lattice distortions as we relaxed the atomic positions.  Generally, an agreement was reached between the computed and measured values\cite{paper}.
\subsection{Electronic density of states}
\label{sec:bands}
 \begin{figure}[h!]
	\begin{center}
		\includegraphics[width=0.4\textwidth]{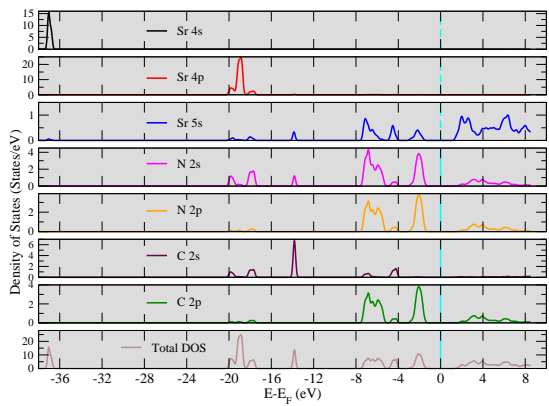}
		\caption{Total and partial density of states of monoclinic Sr$_3$[C$_2$N]$_2$. The DOS is decomposed into the main electron
			states of each component. The vertical line at 0-point (cyan), denotes the fermi-level.}
		\label{fig:bands}
	\end{center}
\end{figure}
The behavior of electrons in a material to external perturbations such as absorption or emission of light can be explained by the energy eigenvalues in the electronic band-structure. Such a response is usually related to an electronic property like the bandgap. A very useful concept in analyzing any band-structure of a material is the density of states as a function of the energy. The calculated total density of states (TDOS) of Sr$_3$[C$_2$N]$_2$ is shown in Fig.~\ref{fig:bands}. We predict monoclinic Sr$_3$[C$_2$N]$_2$ to have an electronic energy gap of about 2.65 eV. Due to the well known GGA bandgap problem, we anticipate that this value is slightly lower than that of the currently missing experimental measurements. Also missing are theoretical bandgaps on Sr$_3$[C$_2$N]$_2$ and this informs why we can not make a comparison of our gap. From the TDOS shown, it is clear that the lower valence bands at about -37 eV are predominantly composed of Sr 4s character. The valence band (VB) and the conduction band (CB) mainly consist of Sr 4p, Sr 5s, N 2s, N 2p, C 2s, and C 2p. Sr 4p and C 2s do not play a role at the top of the valence band and the physics is expected since they are lower in energy. The upper valence band shows strong hybridization between  C 2p N, 2s and N, 2p with very small signatures from Sr 5s. This is a very important observation since the hybridization of N and C states confirms the finding of Clark and co-workers\cite{paper} on C=C=N double bonding. The conduction band consists mainly of C 2p, N 2p N 2s, and Sr 5s. The conduction band harbors these orbitals and the calculated bandwidth is of the order of 7.18 eV. We anticipate that this band characteristic is not rigid and may be tuned to tailor Sr$_3$[C$_2$N]$_2$ for a desired electronic characteristic.

\subsection{Mechanical stability}
\label{sec:elastics}
It is important to note that elastic constants will determine the responses of any solid to external forces. They are characterized by the bulk modulus, Young's modulus, shear modulus, and the Poisson's ratio and play an important role in determining the strength and stability of a material.
\begin{table*}[tb]
	\centering
	\caption{ The calculated 13 elastic constants of Sr$_3$[C$_2$N]$_2$ in GPa.}
	\label{table:1cn}
	\begin{tabular*}{\textwidth}{@{\extracolsep{\fill}}llllllllllllllllll@{}}
		\hline\noalign{\smallskip}
		$C_{11}$& $C_{12}$&$C_{22}$  & $C_{13}$ &$C_{23}$ &$C_{33}$&$C_{44}$&$C_{15}$&$C_{25}$& $C_{35}$&$C_{55}$&$C_{46}$&$C_{66}$\\ 
		\noalign{\smallskip}\hline\noalign{\smallskip}
		103.70&44.22&126.39&45.24&37.81&107.96&31.31&5.15&10.18&-0.31&41.42&0.02&35.00\\
		\noalign{\smallskip}\hline
	\end{tabular*}
\end{table*}
\begin{table*}
	\centering
	\small
	\caption{bulk modulus ($B$), shear modulus ($G$), Young's modulus ($E$), Poisson’s ratio ($\eta$ ) and G/B ratio of Sr$_3$[C$_2$N]$_2$ in GPa, calculated in various schemes; Voigt's, Reuss' and Hill's  approximations.}
	\label{table:3} 
	\begin{tabular*}{\textwidth}{@{\extracolsep{\fill}}lllllllllllllll@{}}
		\hline\noalign{\smallskip}
		\multicolumn{3}{l}{Bulk modulus (GPa)} & \multicolumn{3}{l}{Young's modulus (GPa)} & \multicolumn{3}{l}{Shear modulus (GPa)} & \multicolumn{3}{l}{Poisson's ratio} &\multicolumn{3}{l}{G/B} \\ 
		\noalign{\smallskip}\hline\noalign{\smallskip}
		$B_V$&$B_R$&$B_{H}$&$E_V$&$E_R$ &$E_{H}$&$G_V$ & $G_R$ & $G_{H}$& $\eta_V$ & $\eta_R$ & $\eta_{H}$&$G/B_V$&$G/B_R$&$G/B_H$      \\ 
		\noalign{\smallskip}\hline\noalign{\smallskip}
		65.86&65.05&65.46&90.48&88.32&89.41&35.60&34.67&35.13&0.271&0.274&0.272&0.54&0.53&0.53\\  
		
		\noalign{\smallskip}\hline 
	\end{tabular*}
\end{table*}

In this section, we will introduce the basic formulas of elastic moduli and the mechanical stability criteria for monoclinic Sr$_3$[C$_2$N]$_2$. We investigated the 13 non-zero elastic constants as follows,
\begin{align}
\begin{bmatrix}
C_{11} & C_{12} & C_{13} & 0& C_{15}&0\\
0 & C_{22} & C_{23} & 0& C_{25}&0\\
0 & 0 & C_{33} & 0& C_{35}&0\\
0 & 0& 0 & C_{44}&0&C_{46}\\
0 & 0& 0 & 0&C_{55}&0\\
0 & 0& 0 & 0&0&C_{66}
\end{bmatrix}.
\end{align}

We also calculated the bulk modulus B and shear modulus G using the Voigt approximation\cite{voigt} and the Reuss approximation\cite{reuss}. We then calculated the average of the two to obtain the Hills
approximation\cite{hill}. We used the equations below that relate to the thirteen independent elastic constants to B$_V$ and G$_V$ in Voigt notations, whereas B$_R$ and G$_R$ represent Reuss notations;
\begin{align}
&B_V=\frac{1}{9}\left[C_{11}+C_{22}+C_{33}+2\left(C_{12}+C_{13}+C_{23}\right)\right],
\nonumber\\
&G_V=\frac{1}{15}\left[C_{11}+C_{22}+C_{33}+3\left(C_{44}+C_{55}+C_{66}\right)\right]-
\nonumber\\
&~~~~~~~~\frac{1}{15}\left[\left(C_{12}+C_{13}+C_{23}\right)\right],
\nonumber\\
&B_R=\alpha[ t\left(C_{11}+C_{22}-2C_{12}\right)+u\left(2C_{12}-2C_{11}-C_{23}\right)
\nonumber\\
&~~~~~~~~+v\left(C_{15}-2C_{25}\right)+w\left(2C_{12}+2C_{23}-C_{13}-2C_{22}\right)
\nonumber\\
&~~~~~~~~+2x\left(C_{25}-C_{15}\right)+y]^{-1},
\nonumber\\
&G_R=15\{4[t\left(C_{11}+C_{22}+C_{12}\right)+u\left(C_{11}-C_{12}-C_{23}\right)
\nonumber\\
&~~~~~~~~+v\left(C_{15}+C_{25}\right)+w\left(C_{22}-C_{12}-C_{23}-C_{13}\right)
\nonumber\\
&~~~~~~~~+x\left(C_{15}-C_{25}\right)+y]/\alpha
\nonumber\\
&~~~~~~~~+3[z/\alpha+\left(C_{44}+C_{66}\right)/\left(C_{44}C_{66}-C_{46}^2\right)]\}^{-1}.
\nonumber
\end{align}
Where,
\begin{align}
&t=C_{33}C_{55}-C_{35}^2,
\nonumber\\
&u=C_{23}C_{55}-C_{25}C_{35},
\nonumber\\
&v=C_{13}C_{35}-C_{15}C_{35},
\nonumber\\
&w=C_{13}C_{55}-C_{15}C_{35},
\nonumber\\
&x=C_{C13}C_{25}-C_{15}C_{23},
\nonumber\\
&y=C_{11}(C_{22}C_{55}-C_{25}^2)-C_{12}(C_{12}C_{55}-C_{15}C_{25}),
\nonumber\\
&~~~~+C_{15}(C_{12}C_{25}-C_{15}C_{22})+C_{25}(C_{23}C_{35}-C_{25}C_{33}),
\nonumber\\
&z=C_{11}C_{22}C_{33}-C_{11}C_{23}^2-C_{22}C_{13}^2-C_{33}C_{12}^2
\nonumber\\
&~~~~+2C_{12}C_{13}C_{23},
\nonumber\\
&\alpha=2[C_{15}C_{25}(C_{33}C_{12}-C_{13}C_{23})+C_{15}C_{35}(C_{22}C_13-
\nonumber\\
&~C_{12}C_{23})+C_{25}C_{35}(C_{11}C_{23}-C_{12}C_{13})]-[C_{15}^2(C_{22}C_{33-C_{23}^2})
\nonumber\\
&~~~~+C_{25}^2(C_{11}C_{33}-C_{13}^2)+C_{35}^2(C_{11}C_{22}-C_{12}^2)]+zC_{55}.
\end{align}
It is important to note that the mechanical stability of monoclinic Sr$_3$[C$_2$N]$_2$ is only achieved if the elastic constants satisfy the following necessary and sufficient conditions as prescribed by Born\cite{born}.
\begin{align}
&C_{11} > 0, C_{22} > 0, C_{33} > 0, C_{44} > 0, C_{55} > 0, C_{66} > 0,
\nonumber\\
&\left[C_{11}+C_{22}+C_{33}+2\left(C_{12}+C_{13}+C_{23}\right)\right] > 0,
\nonumber\\
&\left(C_{33}C_{55}-C_{35}^2\right) > 0,
\nonumber\\
&\left(C_{44}C_{66}-C_{46}^2\right) > 0,
\nonumber\\
&\left(C_{22}+C_{33}-2C_{23}\right) > 0,
\nonumber\\
&[C_{22}(C_{33}C_{55}-C_{35}^2)+2C_{23}C_{25}C_{35}-C_{23}^2C_{55}-C_{25}^2C_{33}]>0,
\nonumber\\
&\{2[C_{15}C_{25}(C_{33}C_{12}-C_{13}C_{23})+C_{15}C_{35}(C_{22}C_{13}-C_{12}C_{23})
\nonumber\\
&-C_{25}C_{35}(C_{11}C_{23}-C_{12}C_{13})]-[C_{15}^2(C_{22}C_{33}-C_{23}^2)]
\nonumber\\
&C_{25}^2(C_{11}C_{33}-C_{13}^2)+C_{35}^2(C_{11}C_{22}-C_{12}^2)]+C_{55}z\}>0.
\end{align}
From Table \ref{table:1cn} we noted that Sr$_3$[C$_2$N]$_2$ satisfied all the above tests and is mechanically stable. 
Using the Voigt-Reuss-Hill approximations, we were able to compute B$_H$=0.5(B$_R$+B${_V}$) and equally G$_H$=0.5(G$_R$+G$_{V}$). We obtained Young's modulus $E$ and the Poisson's ratio ($\eta$)  using $E=9BG/(3B+G)$ and $\eta=(3B-2G)/[2(G+3B)]$ respectively. From the calculated elastic constants in Table \ref{table:1cn}, it can be seen that they are too low and we would not expect Sr$_3$[C$_2$N]$_2$ to be employed in the hard industry. The bulk modulus of Sr$_3$[C$_2$N]$_2$ is too small compared to that of diamond (459 GPa\cite{diamond}) indicating that it is a soft material. The comparison to diamond as benchmark for hard materials might be right in this article, but the comparison is somewhat unfair in most calculations. The reason for this is because diamond is dominated by 3D-covalency while in this context, Sr$_3$[C$_2$N]$_2$  by ionic interactions between Sr and C$_2$N. The Young's modulus, which is defined as the ratio between stress and strain, is used to measure stiffness in a solid. A large value of Young's modulus implies stiffness in a material. Sr$_3$[C$_2$N]$_2$ is extremely less stiff if a comparison to diamond is made. We also calculated the value of the Poisson's ratio ($\eta$) in Sr$_3$[C$_2$N]$_2$. This value ranges from -1 to 0.5.  If $\eta=-1$, then the respective material does not change its shape and $\eta=0.5$ implies that the volume does not change. All these happens when an incompressible material is deformed elastically at small strains\cite{elastic}. From Table \ref{table:3}, checking on the Poisson's ratio, we predict that Sr$_3$[C$_2$N]$_2$ is brittle since $\eta<0.33$. We had to confirm this behavior by employing the Pugh's criteria\cite{pugh} according to which a high G/B is associated with brittleness while a low G/B is associated with ductility. In principle, a material is brittle if G/B is greater than 0.5 otherwise the material is ductile. From Table \ref{table:3}, we see that the G/B ratio confirms brittleness in Sr$_3$[C$_2$N]$_2$. Since monoclinic Sr$_3$[C$_2$N]$_2$ is a low symmetry crystal, we caution that calculation of the elastic constants is not unique since it is highly depended on the orientation of the unit cell. At the moment, there is no data on the elastic constants of this crystal and we therefor give a basis for future works to be done.
\subsection{Dynamical stability}
\label{sec:dynamical}
 \begin{figure}[h!]
	\begin{center}
		\includegraphics[width=0.4\textwidth]{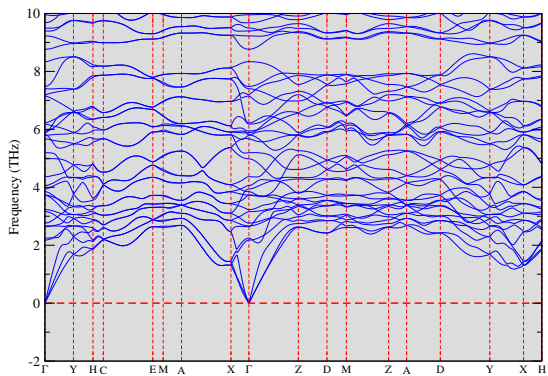}
		\caption{A zoomed phonon band structure of Sr$_3$[C$_2$N]$_2$.}
		\label{fig:ban}
	\end{center}
\end{figure}
The goal here was to compute the phonon band-structure of Sr$_3$[C$_2$N]$_2$ as illustrated in Fig.~\ref{fig:ban}. We computed force constant matrix in real space using the supercell technique based on optimized structures then computed the dynamical matrix for every q-point in reciprocal space. In this case, we observed that all the normal vibration modes had real and finite frequencies thus implying that Sr$_3$[C$_2$N]$_2$ is dynamically stable.
\section{Conclusion}
\label{conc}
We successfully carried out \textit{ab initio} calculations on the mechanical and dynamical stability together with the electronic density of states of Sr$_3$[C$_2$N]$_2$. The goal of this work was to fill part of the information gap on the newly characterized crystal structure. The optimized crystallographic parameters were found to be in accord with experimental results obtained by Clark and co-workers\cite{paper}. The mechanical stability of the new crystal met the stability criteria of a monoclinic system. It was found to be brittle and thus can not be employed where hard materials are required. The dynamical stability was upheld since no imaginary phonon modes were observed in the crystal. The density of states was systemically investigated, and their characteristics were interpreted. It was found out that monoclinic Sr$_3$[C$_2$N]$_2$ has an energy gap of 2.65 eV. We predict that it has a non-rigid gap that can be tuned and make it an ideal candidate for optical applications. Based on these calculated results, we have partly filled the missing information gap on this newly characterized crystal structure. Our major recommendation is that probably if the pseudonitride is combined with suitable cations like in the group of transition metals we may realize subtle physics. It is important to note that at an industrial level, Sr$_3$[C$_2$N]$_2$ can be not be applied as material anyhow due to its reactivity against air and moisture\cite{cooms}. 
\section*{acknowledgement} 
\label{sec:Acknowledgement} 
We thank Rainer Niewa of Universit\"{a}t Stuttgart (Germany), Elkana Rugut of University of the Witwatersrand (South Africa), Samuel Gallego Parra of Universitat Polit\`ecnica de Val\`encia (Spain) and Patrick Ning'i of the Technical University of Kenya (Kenya) for the useful and detailed discussions we had on monoclinic Sr$_3$[C$_2$N]$_2$. GSM acknowledges support from the Kenya Education Network through CMMS mini-grant 2019/2020. The authors also gratefully acknowledge the computer resources, technical expertise, and assistance provided by the Centre for High-Performance Computing (CHPC), Cape Town, South Africa.

\section*{AUTHOR CONTRIBUTIONS} 
\label{sec:contributions}
The research problem was designed by J.S, E.W and C.S then revised by GSM. All the calculations and data analysis were done by J.S and E.W. The manuscript was written by J.S and E.W then revised by  GSM, C.S, S.S and A.O. The project was supervised by C.S and GSM. All the authors approved the manuscript for submission.


\begin{thebibliography}{99}
	\bibliographystyle{apsrev4-1}
	\bibliographystyle{apalike}
\bibitem{james}
George S. Manyali and  James Sifuna, Low compressible $\beta$-BP3N6,
\href{ https://doi.org/10.1063/1.5129268}{ 	AIP Advances 9,12, 125029 (2019)}.
\bibitem{G1}
G. S. Manyali, R. Warmbier, A. Quandt, and J. E. Lowther, \textit{Ab initio} study of elastic properties of super hard and graphitic structures of C$_3$N$_4$,
\href{ https://doi.org/10.1016/j.commatsci.2012.11.039}{ Comput. Mater. Sci. 69 (2013)}.
\bibitem{G2}
G. S. Manyali, R. Warmbier, and A. Quandt, Computational study of the structural, electronic and optical properties of M$_2$N$_2$(NH): M = C, Si, Ge, Sn,
\href{ https://doi.org/10.1016/j.commatsci.2013.07.038}{ Comput. Mater. Sci. 79 (2013)}.
\bibitem{G3}
G. S. Manyali, R. Warmbier, and A. Quandt, First-principles study of Si$_3$N$2$,
\href{https://doi.org/10.1016/j.commatsci.2014.08.051}{ Comput. Mater. Sci. 96 (2015)}.


\bibitem{CN2}
William P. Clark and Rainer Niewa, Synthesis and Characterisation of the Nitridocuprate(I) Nitride Carbodiimide (Sr$_6$N)[CuN$_2$][CN$_2$]$_2$,
\href{ https://doi.org/10.1002/zaac.201900222}{ Z. Anorg. Allg. Chem. 645 (2019)}.
\bibitem{1994}
Ute Berger and Wolfgang Schnick, Syntheses, crystal structures, and vibrational spectroscopic properties of MgCN$_2$, SrCN$_2$, and BaCN$_2$,
\href{ https://doi:org/10.1016/0925-8388(94)90032-9}{ J. Alloys Compd. 206 (1994)}.
\bibitem{2016}
K. B. Sterri, C. Besson, A. Houben, P. Jacobs, M. Hoelzel, R.
Dronskowski, Cr$_2$(NCN)$_3$, a ferromagnetic carbodiimide with an unusual two-step magnetic transition,
\href{ http://dx.doi.org/10.1039/C6NJ02498J}{ New J. Chem.,40  (2016)}.
\bibitem{2001}
Michael Becker and Martin Jansen, Zinc cyanamide, Zn(CN$_2$),
\href{  https://doi.org/10.1107/S0108270101000865}{ Acta Crystallogr., Sect. C, 570  (2001)}.
\bibitem{2004}
R. K. Zheng, Hui Liu, Y. Wang, and X. X. Zhang, Inverted hysteresis in exchange biased Cr$_2$O$_3$
coated CrO$_2$
particles,
\href{  https://doi.org/10.1063/1.1795983}{ Journal of Applied Physics 96, 5370  (2004)}.
\bibitem{paper}
William P. Clark, Andreas K\"ohn, and Rainer Niewa, The Quasi‐Binary Acetonitriletriide Sr$_3$[C$_2$N]$_2$,
\href{  https://DOI: 10.1002/anie.201912831}{ Angew. Chem. Int. Ed. 2020, 59, (2019)}.
\bibitem{msc}
James Sifuna, George S. Manyali, Thomas W. Sakwa and Manasse M. Kitui, Structural and Mechanical Properties of Bulk Scandium Trifluoride Investigated by First-Principles Calculations,
{Journal of Multidisciplinary Engineering Science and Technology 4(2), 6663-6668 (2017)}.
\bibitem{thesis}
James Sifuna, Dental filling materials in the negative and near zero thermal expansion regimes: \textit{Ab initio} study on scandium triflouride,
{Masters dissertation, Faculty of Science, Masinde Muliro University of Science and Technology, Kakamega, 1-64 (2017)}.

\bibitem{siesta}
Jose M. Soler, Emilio Artacho, Julian D. Gale, Alberto Garcia, Javier Junquera, Pablo Ordejon and Daniel Sanchez-Portal, 
Journal of Physics: Condensed Matter
The SIESTA method for \textit{ab initio} order-N materials simulation,
\href{https://doi.org/10.1088/0953-8984/14/11/302}{J. Phys.: Condens. Matter 14 2745 (2002)}.
\bibitem{KB}
Leonard Kleinman and D. M. Bylander, Efficacious Form for Model Pseudopotentials,
\href{https://doi.org/10.1103/PhysRevLett.48.1425}{Phys. Rev. Lett. 48, 1425 {(1982)}}.
\bibitem{TroulMartin}
N. Troullier and José Luís Martins, Efficient pseudopotentials for plane-wave calculations, 
\href{https://doi.org/10.1103/PhysRevB.43.1993}{Phys. Rev. B 43, 1993 {(1991)}}.
\bibitem{monkhorst}
Hendrik J. Monkhorst and James D. Pack, Special points for Brillouin-zone integrations, 
\href{https://doi.org/10.1103/PhysRevB.13.5188}{Phys. Rev. B 13, 5188  {(1976)}}.
\bibitem{soler_kpts}
Juana Moreno and José M. Soler, Optimal meshes for integrals in real- and reciprocal-space unit cells,
\href{https://doi.org/10.1103/PhysRevB.45.13891}{Phys. Rev. B 45, 13891 {(1992)}}.
\bibitem{pbe}
John P. Perdew, Kieron Burke, and Matthias Ernzerhof, Generalized Gradient Approximation Made Simple, 
\href{https://doi.org/10.1103/PhysRevLett.77.3865}{Phys. Rev. Lett. 77, 3865 {(1996)}}.


\bibitem{munaghan}
F. D. Murnaghan, The Compressibility of Media under Extreme Pressures, 
\href{https://doi.org/10.1073/pnas.30.9.244}{Proc. Natl. Acad. Sci. U. S. A. 30, 244 (1944).}



\bibitem{voigt}
W. Voigt, Lehrbuch der Kristallphysik,
{Teubner, Leipzig {(1928)}}.
\bibitem{reuss}
A. Reuss, Berechnung der Fließgrenze von Mischkristallen auf Grund der Plastizitätsbedingung für Einkristalle,
\href{https://doi.org/10.1002/zamm.19290090104}{Z. Angew. Math. Mech. 9 {(1929)}}.
\bibitem{hill}
R. Hill, 
Proceedings of the Physical Society. Section A
The Elastic Behaviour of a Crystalline Aggregate,  
\href{https://doi.org/10.1088/0370-1298/65/5/307}{Proc. Phys. Soc., Sect. A 65 {(1953)}}.
\bibitem{creteria}
Zhi-jian Wu, Er-jun Zhao, Hong-ping Xiang, Xian-feng Hao, Xiao-juan Liu, and Jian Meng, Crystal structures and elastic properties of superhard IrN$_2$ and IrN$_3$ from first principles,
\href{https://doi.org/10.1103/PhysRevB.76.054115}{Phys. Rev. B 76, 054115 {(2007)}}.

\bibitem{born}
M. Born and K. Huang, Dynamical theory of crystal lattices,
\href{http://cds.cern.ch/record/224197}{Oxford : Clarendon Press {(1954)}}.

\bibitem{diamond}
M.Hebbache, First-principles calculations of the bulk modulus of diamond,
\href{https://doi.org/10.1016/S0038-1098(99)00122-2}{Solid State Communications. 110,10 {(1999)}}.

\bibitem{elastic}
N. Wang, W.Y. Yu, B.-Y. Tang, L.M. Peng, and W.J, Structural and mechanical properties of Mg$_17$Al$_12$ and Mg$_2$4Y$_5$ from first-principles calculations,
\href{https://doi.org/10.1088/0022-3727/41/19/195408}{J. Phys. D: Appl. Phys. 41 195408 {(2008)}}.

\bibitem{pugh}
S. Pugh, XCII. Relations between the elastic moduli and the plastic properties of polycrystalline pure metals,
\href{https://doi.org/10.1080/14786440808520496}{Philosophical Magazine, vol. 45, 367 {(1954)}}.
\bibitem{cooms}
Private communications with Prof. Rainer Niewa,
\href{https://www.iac.uni-stuttgart.de/en/research/akniewa/}{on 05-01-2020}.
\end {thebibliography}

\end{document}